\documentclass[fleqn,usenatbib]{mnras}

\usepackage{newtxtext,newtxmath}

\usepackage[T1]{fontenc}

\DeclareRobustCommand{\VAN}[3]{#2}
\let\VANthebibliography\thebibliography
\def\thebibliography{\DeclareRobustCommand{\VAN}[3]{##3}\VANthebibliography}

\usepackage{graphicx}	
\usepackage{amsmath}	
\usepackage{pdflscape}	

\title[Understanding the origin of early-type dwarfs]{Understanding the origin of early-type dwarfs: The spectrophotometric study of CGCG014-074}
\author[N. Guevara et al.]{
N. Guevara,$^{1,2}$\thanks{E-mail: guevaran@fcaglp.unlp.edu.ar}
C. G. Escudero,$^{1,2}$
F. R. Faifer$^{1,2}$
\\
$^{1}$Facultad de Ciencias Astron\'omicas y Geof\'isicas of the Universidad Nacional de La Plata,Paseo del Bosque S/N, B1900FWA, La Plata, Argentina \\
$^{2}$Instituto de Astrofísica de La Plata (Conicet), Paseo del Bosque S/N, B1900FWA, La Plata, Argentina 
}

\date{Accepted XXX. Received YYY; in original form ZZZ}

\pubyear{2023}

\begin{document}
\label{firstpage}
\pagerange{\pageref{firstpage}--\pageref{lastpage}}
\maketitle

\begin{abstract}
Early-type dwarf galaxies constitute a prevalent population in the central regions of rich groups and clusters in the local Universe. These low-luminosity and low-mass stellar systems play a fundamental role in the assembly of the luminous galaxies observed today, according to the $\Lambda$CDM hierarchical theory. The origin of early-type dwarfs has been linked to the transformation of disk galaxies interacting with the intracluster medium, especially in dense environments. However, the existence of low-luminosity early-type galaxies in low-density environments presents a challenge to this scenario.
This study presents a comprehensive photometric and spectroscopic analysis of the early-type dwarf galaxy CGCG014-074 using deep GEMINI+GMOS data, focusing on its peculiarities and evolutionary implications. CGCG014-074 exhibits distinct features, including a rotating inner disk, an extended stellar formation with a quiescent phase since about 2 Gyr ago, and the presence of boxy isophotes. 
From the kinematic analysis, we confirm CGCG014-074 as a nucleated early-type dwarf galaxy with embedded disk. The study of its stellar population parameters using different methods provides significant insights into the galaxy's evolutionary history. These results show an old and metal-poor nucleus ($\sim 9.3$ Gyr and $\mathrm{[Z/H]}\sim-0.84$ dex), while the stellar disk is younger ($\sim4.4$ Gyr) with a higher metallicity ($\mathrm{[Z/H]}\sim-0.40$ dex).
These distinctive features collectively position CGCG014-074 as a likely building block galaxy that has evolved passively throughout its history.

\end{abstract}




\section{Introduction}

Dwarf galaxies constitute a class of low-luminosity ($M_B > -18$ mag) and low-mass ($M < 10^9 M_\odot$) stellar systems. Among these, early-type dwarf galaxies 
stand out as the dominant galaxy type within nearby clusters and groups \citep{Ferguson1994}. According to the $\Lambda$CDM hierarchical theory \citep[such as][]{White1991,Valcke2008,Shen2014}, dwarf galaxies serve as the fundamental constituents in the assembly of the luminous galaxies observed today.

The formation scenarios and theories for these objects focus on the transformation of late-type galaxies into their early-type counterparts through different processes involving interactions with their cluster/group and the environment. The galactic cold gas depletion induced by the hot intracluster medium's ram pressure \citep{Gunn1972}, tidal mass loss and kinematic heating caused by tidal shocks \citep{Moore1998}, and structural alterations from tidal interactions with massive cluster members are all crucial factors in this transformation. These effects are particularly pronounced in low-mass galaxies due to their shallow gravitational potentials. Furthermore, the morphology-density relation underscores the affinity of early-type galaxies for high-density regions \citep{Dressler1980,Binggeli1990}. Thus, the combined impact of ram pressure, environmental quenching \citep{Peng2010}, and tidal interactions leads to gas removal, suppression of star formation, and kinematic heating.

Supporting evidence for this theory emerges from observations of dwarf galaxies retaining vestiges of their late-type history, such as bars or residual spiral structures \citep[e.g.,][]{Lisker2006,Janz2014,SeoAnn2022}. Additionally, the study of intermediate luminosity jellyfish galaxies provides insight into the active interplay of these mechanisms \citep{Poggianti2017}. The presence of fast-rotating dE-like galaxies also contributes to this framework \citep{Toloba2015,Penny2016,Scott2020}. However, it remains unclear whether these processes are relevant in low-density environments, such as the field or poor groups \citep{Annibali2011}. Hence, a comprehensive analysis of the morphological characteristics of dwarf galaxies becomes essential for unravelling their underlying dynamics and assembly histories.

In this paper, we present a photometric and spectroscopic analysis of CGCG014-074 \citep[$M_B\approx-$15.5;][]{Grogin1998}, a completely unexplored early-type dwarf galaxy originally catalogued as a dwarf lenticular \citep[dS0;][]{deVaucouleurs1991}, located in the vicinity of NGC\,4546 \citep[$M_V=-20.18$;][]{Escudero2020}, a massive lenticular galaxy classified as a field galaxy \citep[$\log(\rho)=-1.14$ Mpc$^{-3}$;][]{Cappellari2011}. 
The observational properties of both galaxies are summarized in Table \ref{tab:observ_prop}. By considering the distance modulus for the group of $(m-M)=30.75\pm0.12$ mag ($14.1\pm1.0$ Mpc), the spatial scale corresponds to 0.067 kpc/arcsec.
Given their location in a low-density environment, studying this galactic pair offers an exceptional case to study the formation and evolution processes of dwarf and S0 galaxies under such conditions. Our main goal is to characterize the evolutionary history of CGCG014-074 by examining its stellar populations globally and across spatially resolved regions.
From this analysis, we hope to determine if the morphology of this early-type dwarf galaxy reflects gradual formation and evolution as a building block, or if significant mergers and/or interactions have influenced it. The first scenario, according to cosmological hierarchical formation models, predicts an extended star formation history (SFH), leading to enriched stellar populations over time until their cold gas is depleted, while the latter case suggests a radial gradient of age and metallicity, with a younger, metal-rich central region \citep[see e.g.,][]{Koleva2009,Ann2024}.

The paper is organised as follows. Section \ref{sec:obs_data} introduces the data and its reduction procedure, while Section \ref{sec:analysis} provides a comprehensive account of the photometric and spectroscopic analysis performed. Finally, Section \ref{sec:summary} discusses the results obtained in this study.

\begin{table}
	\centering
    \defcitealias{NedIpac}{NASA/IPAC Extragalactic Database}
	\caption{Astrophysical properties of CGCG014-074 and NGC4546. References: [1] \citetalias{NedIpac}; [2] \citet{Paturel2005}; [3] \citet{Colless2003}; [4] \citet{Cappellari2011}; [5] \citet{Cappellari2013}; [6] \citet{Escudero2020}; [7] \citet{Norris2011}.}
     \defcitealias{NedIpac}{$1$}
     \defcitealias{Paturel2005}{$2$}
     \defcitealias{Colless2003}{$3$}
     \defcitealias{Cappellari2011}{$4$}
     \defcitealias{Cappellari2013}{$5$}
     \defcitealias{Escudero2020}{$6$}
     \defcitealias{Norris2011}{$7$}
	\label{tab:observ_prop}
	\begin{tabular}{ccccc} 
		\hline
        Property & CGCG014-074 & NGC 4546 & unit & references\\
		\hline
		$\mathbf{\alpha}$             &  12:35:50.95         &  12:35:29.5          &  h:m:s (J2000) & \citepalias{NedIpac} \\
        $\mathbf{\delta}$             &  $-$03:45:58.5       &  $-$03:47:35.5       &  d:m:s (J2000) & \citepalias{NedIpac} \\
        $l$                           &  295:23:32.4         &  295:13:38.5         &  d:m:s         & \citepalias{NedIpac} \\
        $b$                           &  58:52:37.6          &  58:50:23.3          &  d:m:s         & \citepalias{NedIpac} \\
        Type                          &  dS0 edge-on         &  SB0$^{-}$(s)        &  --            & \citepalias{NedIpac}  \\
        ${I}_{\mathbf T}^{\mathbf 0}$ &  14.17$\pm$0.15      &  9.33$\pm$0.36       &  mag           & \citepalias{Paturel2005}  \\
        $V_\mathrm{hel}$              &   998$\pm$54         &  1057$\pm$5          &  km/s          & \citepalias{Colless2003,Cappellari2011}\\
        $\sigma_e$                    &  --                  &  188                 &  km/s          & \citepalias{Cappellari2013} \\
        $R_\mathrm{eff}$              &  --                  &  22.23               &  arcsec        & \citepalias{Cappellari2013} \\        
        $(m-M)_0$                     &  --                  &  30.75$\pm$0.12      &  mag           & \citepalias{Escudero2020} \\
        Dist.                         &  --                  &  14.1$\pm$1.0        &  Mpc           & \citepalias{Escudero2020} \\
        $M_\star$                     &  --                  &  2.7$\times10^{10}$  & M$_\odot$      & \citepalias{Norris2011} \\
        \hline
	\end{tabular}
\end{table}

\section{Observational data}
\label{data}
\begin{figure}
    \centering
    \includegraphics[]{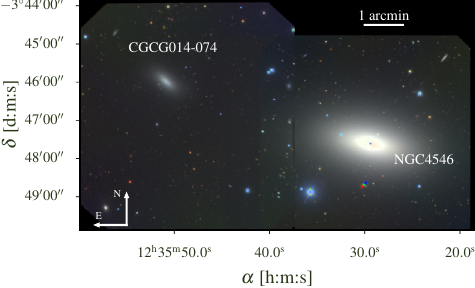}
    \caption{Mosaic of GEMINI-GMOS images showing the position in the sky of the \textbf{dwarf} galaxy CGCG014-074 and the S0 galaxy NGC\,4546. These galaxies are located at a distance of $\sim$14 Mpc, with a projected spatial separation between them of $\sim$ 22\,kpc.}
    \label{fig:map-galaxies}
\end{figure}
\label{sec:obs_data}
The acquisition of photometric data was carried out under the programme GS-2014A-Q-30 (PI: Escudero, C.), using the capabilities of GEMINI/GMOS instrument in image mode. The observations were conducted under exceptional seeing conditions, achieving an FWHM between $0.46-0.67$ arcsec. The dataset comprises exposures in the $g'$, $r'$, $i'$, and $z'$ filters \citep{Fukugita1995}, each consisting of four exposures of $100$ sec for $g'$, $r'$, and $i'$, and four exposures of $290$ sec for the $z'$ filter. 

The observations were reduced using specific GEMINI/GMOS routines within the \textsc{iraf}\footnote{IRAF is distributed by the National Optical Astronomical Observatories, which are operated by the Association of Universities for Research in Astronomy, Inc., under cooperative agreement with the National Science Foundation} software (version 2.16), such as \textsc{gprepare}, \textsc{gbias}, \textsc{giflat}, \textsc{gireduce}, and \textsc{gmosaic}. Baseline calibration was implemented using bias and flat-field images obtained from the Gemini Observatory Archive (GOA), which were essential for correcting the raw data. 
In particular, the $i'$ and $z'$ frames exhibited night sky fringing caused by thin film interference effects within the CCD detectors. To mitigate this effect, blank sky frames were used to subtract the fringing pattern. 
These calibration images were downloaded from the GOA, combined with the \textsc{gifringe} task, and applied to the science frames using the \textsc{girmfringe} task. Finally, the \textsc{iraf} task \textsc{imcoadd} was used for the co-addition of the reduced frames in each filter, obtaining the final $g'$, $r'$, $i'$, and $z'$ images. Figure \ref{fig:map-galaxies} shows the GMOS mosaic with the position of CGCG014-074 and NGC\,4546.

The spectroscopic observations, on the other hand, were obtained with the GMOS instrument in long-slit mode, under the programme GS-2020A-Q-130 (PI: Escudero, C.). A total of nine exposures of $1540$ sec were taken using the B1200 grating and a slit of $1$ arcsec width slit aligned along the major axis of the galaxy. Taking into account the gap between the CCDs and the slit bridges, the data acquisition was centred at three different wavelengths, $530, 540$ and $550$ nm, and with a spatial offset of $7$ arcsec. This instrumental configuration yields a dispersion of $0.26$\,\AA/pixel with a spectral resolution of FWHM$\sim1.9$\AA\, measured at several sky lines, and covering a wavelength range of $4700$-$6200$\,\AA. 

The subsequent reduction process was performed using the tasks from the GEMINI/GMOS \textsc{iraf} package, run within the \textsc{pyraf} environment (version 2.1.15 for Python 2.7). Bias calibrations were downloaded from the GOA and used in the reduction process along with the flat-field and arc calibrations observed on the same night as the science frames. The latter were corrected, rectified and calibrated in wavelength using tasks such as \textsc{gbias}, \textsc{gqecorr}, \textsc{gsflat}, \textsc{greduce}, \textsc{gswavelenght} and \textsc{gstransform}. Cosmic rays were removed from the images using the \textit{Laplacian Cosmic Ray Identification} \citep{vanDokkum2001} routine via the \textsc{gemcrspec} task. To obtain a single 2D image of the spectrum, the nine individual science frames were combined using the \textsc{lscombine} task, taking into account both the spectral and spatial offset. The standard star CD-329927 underwent the same reduction procedure, and its sensibility function was derived using the \textsc{gsstandard} task.
Flux calibration and heliocentric correction of the 1D science spectra were performed using the sensitivity function and the \textsc{calibrate} and \textsc{dopcor} tasks, respectively.

\section{Analysis}
\label{sec:analysis}
\subsection{Surface Brightness Profiles}
\label{profiles}
The surface brightness distribution of the dwarf galaxy was studied using the \textsc{ellipse} task \citep{Jedrzejewski1987} from the \textsc{iraf} software. The isophotal parameters (ellipticity ($\varepsilon$), position angle (PA), and Fourier coefficients $A_4, B_4$) were allowed to vary as a function of the equivalent radius ($r_\mathrm{eq}= a \sqrt{1-\varepsilon}$; where $a$ is the semi-major axis of the ellipses).

\textsc{ellipse} was run interactively on the images, with appropriate masking of the light from NGC\,4546 and bright objects in the field before fitting. During the fitting process, the ellipses' centre, $\varepsilon$, and PA were allowed to vary freely. However, as the iterations extended towards the outer regions of the galaxy where the signal-to-noise ratio (SNR) decreases, the aforementioned parameters were fixed to achieve convergence in the fit. In this case, the edge of the images was reached, corresponding to the equivalent radius of 60 arcsec.

Finally, the models obtained in each filter were calibrated to the standard system using the following expression:
\begin{equation}\label{eq:elipse_calib}
\mu(r)= C_0 - 2.5\,\log_{10}\left(\frac{I(r)}{t\,E^{2}}\right) - K \left( X -1 \right),
\end{equation}
where $\mu(r)$ represents the surface brightness at the equivalent radius $r$ in magnitudes per square arcsec, $C_0$ denotes the constant for transformation to the standard photometric system derived from \citet{Escudero2020}{, and it contains the galactic extinction and the point-zero correction}, $I(r)$ corresponds to the mean intensity of the isophote in ADU per square pixel, $t$ is the exposure time of the data (see Section \ref{sec:obs_data}), $E$ is the GMOS detector scale in arcsec per pixel ($0.146$ arcsec/pixel), $K$ is the mean atmospheric extinction and $X$ the airmass of the observations.

To validate and refine sky-level estimations in the profiles, images in $g$, $r$, $i$ and $z$ bands were downloaded from the DESI Legacy Imaging Surveys\footnote{http://legacysurvey.org/} \citep{Dey2019}. These images, although shallower photometrically than the GMOS images, cover a larger area around the galaxy. Again, \textsc{ellipse} was used on this dataset, following the same procedure as on the GMOS images, but reaching a larger galactocentric radius ($r \sim 80$ arcsec; 5.6 kpc). Subsequently, the profiles obtained from the Legacy images were calibrated using Equation \ref{eq:elipse_calib}, considering a pixel scale of $0.27$ arcsec/pixel, and $C_0=22.5$ mag, for all photometric bands. Finally, the calibrated profiles from GMOS and the corresponding Legacy images were compared to unveil any subtle signal differences between them, mainly due to the effect of the background sky value considered during the isophotal fit. In this way, the calibrated surface brightness profiles of CGCG014-074 were obtained in the four filters.

\begin{figure}
	\includegraphics{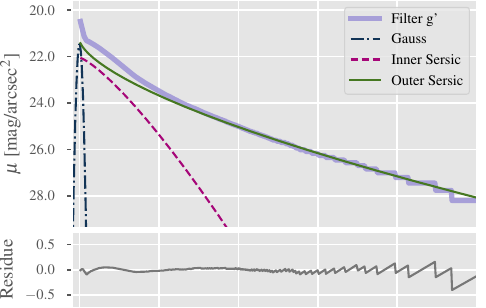}
    \includegraphics{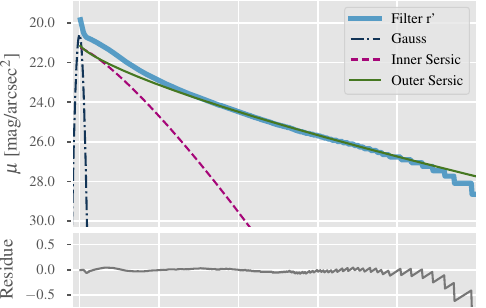}
    \includegraphics{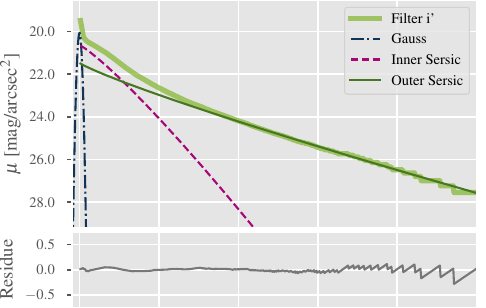}
    \includegraphics{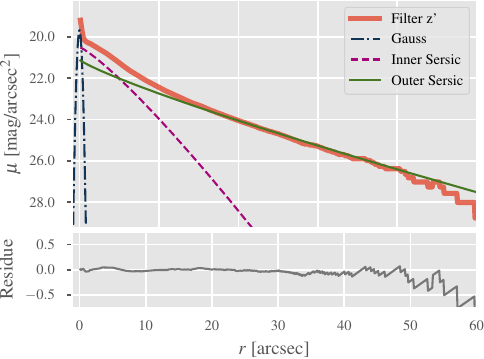}
    \caption{Surface brightness profiles for the different filters and their parametric profile fitting. Dark blue dashed and dotted lines represent Gaussian profiles, magenta dashed lines show the innermost Sérsic profile, and green solid lines show the outermost Sérsic profile. The lower panels display the residuals of the combined fit.}
    \label{fig:ajustes_perfiles}
\end{figure}

To characterise and estimate the structural parameters of CGCG014-074, a Sérsic \citep{Sersic1968} function was initially fitted in the form: 
\begin{equation}\label{eq:sersic}
\mu(r)= \mu_\mathrm{eff} + \left({1.086\,b_n}\right)\left[\left(\frac{r}{R_\mathrm{eff}}\right)^{(1/n)}-1\right],
\end{equation}
where $R_\mathrm{eff}$ is the effective radius of the galaxy, $\mu_\mathrm{eff}$ is the surface brightness at $R_\mathrm{eff}$, $n$ represents the Sérsic index, and $b_n$ is a parameter associated with $n$ that follows the approximate expression $b_n=1.9992,n-0.3271$ for $0.5<n<10$ \citep{Graham2008} to ensure that half of the total flux is within $R_\mathrm{eff}$.

\begin{table} 
	\centering
	\caption{Parameters obtained by fitting a Sérsic profile to the surface brightness profiles, and the estimated total magnitudes in each filter.}
	\label{tab:ajustes_perfil_sersic}
    \resizebox{\columnwidth}{!}{
	\begin{tabular}{lccccc} 
		\hline
		  Filter  &  $\mu_\mathrm{eff}$ &  $R_\mathrm{eff}$    &  $n$  &  $m_0$  & $M_0$  \\
           & (mag/arcsec$^2$)  &  arcsec (kpc)  &       &  (mag)  & (mag) \\
		\hline
		$g'$    &  23.66$\pm$0.01  &  13.2$\pm$0.04\,(0.88)  &  1.69$\pm$0.01  &  15.08$\pm$0.11  &  -15.65$\pm$0.17 \\
$r'$    &  23.09$\pm$0.01  &  13.2$\pm$0.04\,(0.88)  &  1.71$\pm$0.01  &  14.50$\pm$0.10  &  -16.23$\pm$0.17 \\
$i'$    &  22.85$\pm$0.01  &  13.6$\pm$0.04\,(0.91)  &  1.73$\pm$0.01  &  14.22$\pm$0.11  &  -16.51$\pm$0.17 \\
$z'$    &  22.58$\pm$0.01  &  13.1$\pm$0.04\,(0.87)  &  1.69$\pm$0.01  &  14.05$\pm$0.10  &  -16.68$\pm$0.17 \\
		\hline
	\end{tabular}
    }
\end{table}

The parameters obtained from these fits are listed in Table \ref{tab:ajustes_perfil_sersic}, where the integrated apparent (and absolute) magnitudes of the galaxy were also obtained using the following expression: 
\begin{equation}\label{eq:mag_total}
  {m}_0 = \mu_\mathrm{eff}-1.995450-5~\log(R_\mathrm{eff})-1.0857~b_n-2.5~\log\left[b_n^{-2n}~n\Gamma(2n)\right] 
\end{equation}
being $\Gamma(2n)$ the gamma function. As observed in the mentioned Table, the values of the effective radii and Sérsic indices obtained in the different bands are similar, being typical values obtained for early-type dwarf galaxies \citep{Aguerri2005,Janz2008,Janz2014}. This characteristic would indicate that strong colour gradients are not expected in the galaxy.

These single-component fits give reasonable results, but when the residuals are examined in detail, particularly towards the innermost and outermost regions of the profiles ($r_\mathrm{eq}<10$ and $r_\mathrm{eq}>55$ arcsec), they show values higher than $\pm0 .2$ mag/arcsec$^2$. This suggests that a single component is not sufficient to accurately reproduce the galaxy's light distribution. Therefore, the next step was to fit multiple functions to the profiles. This was done following the procedure of \citet{Huang2013}, where the best model is the one that contains a minimum number of components with reasonable parameters that describe visibly different structures.

First, a Sérsic profile was considered to model the extended component of the galaxy. The fit was done by leaving the model parameters free for $r_\mathrm{eq}>10$ arcsec to avoid the excess light shown by the profiles in the central region as well as the effect of seeing. This model was then subtracted from the profile to remove the extended component of the galaxy, and the residual was observed to confirm that the model was correct. A second Sérsic profile was then used to model the excess light between $0.6<r_\mathrm{eq}<10$ arcsec. As mentioned above, this model was subtracted from the profile by verifying the residual obtained. Finally, to represent the inner region of the galaxy, a Gaussian profile was fitted in the range $r_\mathrm{eq}<0.6$ arcsec. From these initial values obtained for each function, they were fitted simultaneously throughout the galactocentric range to improve and obtain the final parameters. 

\begin{table*} 
	\caption{Best parameters obtained for each fitted function.}
	\label{tab:ajustes_perfiles}
	\begin{tabular}{lcccccc} 
		\hline
		    Filter  & Model &  $\mu_\mathrm{eff}$        & R$_\mathrm{eff}$  &  $n$  & $\sigma$   &  $rms$ \\
           &        &   (mag\:arcsec$^{-2}$) & (arcsec-{pc})  &       & (arcsec-{pc})  &  \\
		\hline
$g'$    & Gaussian  & $21.46\pm0.15$  &  --                    & --             & $0.26\pm0.03-{17.5\pm 1.8}$  & $0.04$\\
        & Sérsic     & $23.38\pm0.15$  &  ~$5.91\pm0.11-{393.9 \pm 7.8}$   & $0.78\pm0.09$  &  --  &\\
        & Sérsic     & $24.28\pm0.11$  &  $16.98\pm0.60-{1133.0 \pm 42.8}$   & $1.54\pm0.09$  &  --  & \\
\hline
\rule{0pt}{1.05em}%
$r'$    & Gaussian  & $20.59\pm0.07$  &  --                   & --             & $0.27\pm0.01-{18.2 \pm 1.0}$  & $0.03$  \\
        & Sérsic     & $22.61\pm0.07$  &  ~$5.75\pm0.10-{383.9 \pm 7.7}$  & $0.79\pm0.04$ &  --  &  \\
        & Sérsic     & $23.69\pm0.07$  &  $16.66\pm0.43-{1126.5 \pm 26.6}$ & $1.36\pm0.05$  &  --  &  \\
\hline
\rule{0pt}{1.05em}%
$i'$    & Gaussian  & $20.06\pm0.08$  &  --                  & --             & $0.24\pm0.01-{16.3 \pm 0.9}$  & $0.04$\\
        & Sérsic     & $22.15\pm0.04$  &  ~$5.93\pm0.15-{398.4 \pm 10.2}$  & $0.86\pm0.03$ &  --  & \\
        & Sérsic     & $23.61\pm0.07$  &  $18.00\pm0.47-{1206.9 \pm 31.6}$ & $1.15\pm0.05$  &  --  & \\
\hline
\rule{0pt}{1.05em}%
$z'$    & Gaussian  & $19.67\pm0.10$  &  --                   & --             & $0.23\pm0.01-{14.7 \pm 1.2}$  & $0.05$\\
        & Sérsic     & $21.96\pm0.05$  &  ~$5.82\pm0.17-{480.7 \pm 20.0}$   & $0.84\pm0.04$ &  --  &  \\
        & Sérsic     & $23.31\pm0.09$  &  $17.02\pm0.55-{1411.2 \pm 55.1}$  & $1.18\pm0.06$  &  --  &  \\
		\hline
	\end{tabular}
    
\end{table*}

Figure \ref{fig:ajustes_perfiles} shows the best model obtained for each filter, composed of three functions considered, while Table \ref{tab:ajustes_perfiles} lists the values obtained for each component. It can be seen that the inner component, described by a Sérsic profile, has an index value of $n\approx0.8$. This value agrees with what is expected for early-type dwarf galaxies with disk properties \citep{Allen2006, Graham2012, SeoAnn2022}.

\subsection{Maps and Colour Profiles}\label{sec:colourmaps}

To obtain the profiles and colour maps of CGCG014-074, we initially aligned all the photometric images using the $g'$ filter image as a reference, using the \textsc{iraf} tasks \textsc{geomap} and \textsc{geotran}. Then, since the $g'$ filter frame is the lowest quality (FWHM$=$0.67 arcsec), the $r',i'$ and $z'$ images were degraded to this value to homogenise the dataset and avoid substructures in the colour maps and profiles caused by differing FWHM values.
This involved degrading the images using the \textsc{gauss} task within \textsc{iraf}, which convolves the original science image with a Gaussian kernel defined by its standard deviation ($\sigma$).
To denoise the colour map, we applied a smoothing process to these standardised images using the \textsc{boxcar} task with a rectangular kernel of dimensions $5\times5$ pixels, similar to the seeing value. Finally, the \textsc{imcalc} task facilitated the generation of colour maps by calculating the differences between the images. Obtaining these colour maps can provide relevant information about different stellar components and/or colour substructures related to the presence of dust regions and/or merger events.

\begin{figure*}
    \centering
    \includegraphics[width=.3\linewidth ]{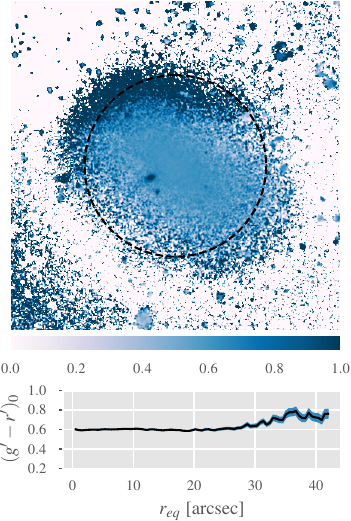}
    \includegraphics[width=.3\linewidth ]{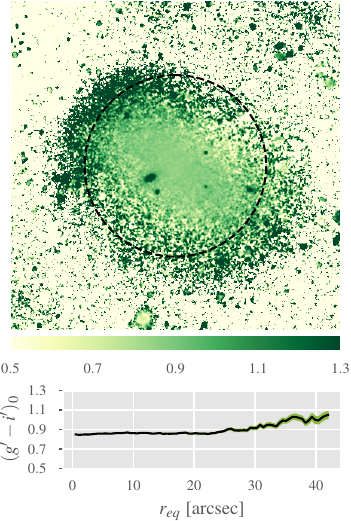}
    \includegraphics[width=.3\linewidth ]{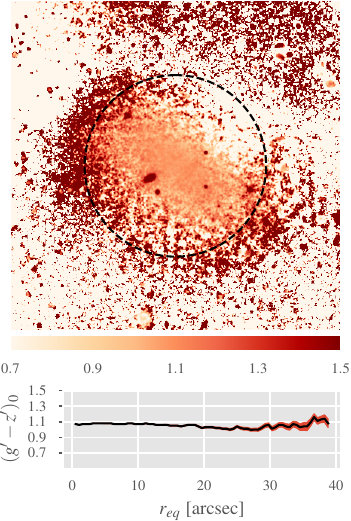}
    \caption{Colour maps and colour profiles $(g'-r')_0$ (left panel), $(g'-i')_0$ (central panel), and $(g'-z')_0$ (right panel). The black dashed line indicates a circle of $r_\mathrm{eq}=40$ arcsec where the colour profiles were obtained. }
    \label{fig:color_maps_profiles}
\end{figure*}

The \textsc{ellipse} task is used to extract the colour profiles from the generated colour maps. Fixed parameters for centre position, ellipticity and position angle were adopted, based on average values ($\langle \epsilon \rangle = 0.47$ and $\langle AP \rangle = 37^\circ$) derived from calibrated surface brightness profiles (as described in Section \ref{profiles}). These parameters allowed us to construct corresponding colour profiles for the colour indices $(g'-r')_0$, $(g'-i')_0$, and $(g'-z')_0$. 
Figure \ref{fig:color_maps_profiles} presents the colour maps and profiles of CGCG014-074. These profiles extend up to $r_\mathrm{eq} \sim 40$ arcsec (indicated by a black circle with a dashed line in the colour maps). Beyond this radius, the considered sky values begin to significantly influence the colour maps.

As can be seen, the colour profiles remain relatively flat within $r_\mathrm{eq}<30$ arcsec ($r_\mathrm{eq}<2$ kpc), with mean colours of $\langle(g'-r')_0\rangle=0. 6$, $\langle(g'-i')_0\rangle=0.86$, and $\langle(g'-z')_0\rangle=1.07$ mag, with dispersions of $\sigma_{g'-r}=0.01$, $\sigma_{g'-i}=0.02$, and $\sigma_{g'-z}=0.03$ mag. Moreover, the colour maps do not show any discernible signs of dust presence or distinct colour substructures.

\subsection{Isophotal Analysis}

The results of the fitting process performed by the \textsc{ellipse} task on the photometric images provide a basis for analyzing the variations of the isophotal parameters ($\varepsilon$, PA, cosine Fourier coefficient $B_4$) relative to the equivalent radius.

\begin{figure}
    \includegraphics[scale=.95]{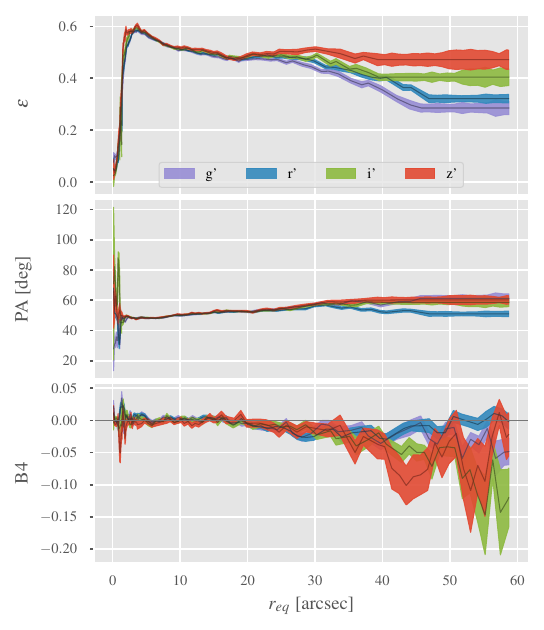} 
    \caption{Isophotal parameters as a function of equivalent radii, result of the photometric analysis performed with \textsc{ellipse}, for all filters. The top panel shows the variation of the ellipticity ($\varepsilon$), the middle panel the variation of the position angle (PA) and the bottom panel the variation of the Fourier harmonic amplitude $B_4$.}
    \label{fig:param_ellipse}
\end{figure}

Figure \ref{fig:param_ellipse} shows the variation of the aforementioned parameters in different photometric bands. 
In the innermost region ($r_\mathrm{eq}<3$ arcsec; $0.2$ kpc), the ellipticity shows significant variation, ranging from $\varepsilon \sim 0.1$ to $0.6$. This large shift is due to the presence of the nucleus of the dwarf galaxy. Beyond $r_\mathrm{eq}>3$ arcsec, the ellipticity begins to decrease smoothly until $r_\mathrm{eq}\sim18$ arcsec ($1.2$ kpc), remaining at an approximately constant value ($\varepsilon \sim0.48$) until $r_\mathrm{eq}\sim30$ arcsec ($1.8$ kpc). 

On the other hand, the position angle (PA), measured counterclockwise from north to east, shows a similar behaviour to the ellipticity in the inner region ($r_\mathrm{eq}<3$ arcsec; $0.2$ kpc), changing its value rapidly. From $r_\mathrm{eq}=3$ to $60$ arcsec, the PA gradually increases, varying by about $13$ degrees between the two ends, reaching a value of 60 degrees in the outermost region of the galaxy. It is interesting to note that the fitted isophotes begin to rotate smoothly, pointing in the direction of the companion galaxy NGC\,4546. 

Finally, regarding the Fourier coefficient $B_4$, it takes values $B_4>0$ for $r_\mathrm{eq}<18$ arcsec ($1.2$ kpc), indicating the presence of disk-like isophotes. These types of isophotes are found in the galactocentric region where $\varepsilon$ shows a smooth change, pointing to the presence of an inner stellar disk. 
Then, as we move towards larger radii ($r_\mathrm{eq}>18$ arcsec), the parameter shows values $B_4<0$, indicating boxy-like isophotes.

These characteristics exhibited by CGCG014-074 resemble those of the rectangular-shaped galaxy LEDA\,074886, studied by \citet{Graham2012}, which also shows an edge-on stellar disk and notable boxy-like isophotes.

\subsection{Kinematics}\label{sec:kinematics}

To study the kinematics of CGCG014-074, 1D spectra at different galactocentric radii were extracted from the 2D spectroscopic image (see Section \ref{sec:obs_data}). This was done using an iterative IDL code (version 7.1), which varies the aperture of a given extraction around certain galactocentric radii to obtain an SNR above a user-defined value. In this work, a criterion of $SNR > 20$ per \AA\ (measured at $4700$\AA) was chosen to ensure a robust analysis of the kinematics and stellar population history of the galaxy. However, for regions beyond $15$ arcsec ($r_g > 15$ arcsec; 1 kpc) in the 2D spectrum, this condition was relaxed to $SNR > 5$ per \AA\ to allow at least a meaningful kinematic analysis.

From these spectra, we determined the radial velocity ($V_\mathrm{rad}$) and the velocity dispersion ($\sigma_*$) using the full spectral fitting technique implemented in the \textsc{pPXF} algorithm \citep{Cappellari2004, Cappellari2017}. In this work, we used the stellar population synthesis models from \citet{Maraston2011}, specifically the single stellar population (SSP) models based on the ELODIE stellar library. This preference was driven by the spectral range coverage, aligning closely with our observations ($3900-6800$ \AA), albeit at a slightly higher resolution (FWHM$=0.55$ \AA) compared to the CGCG014-074 spectra (FWHM$=1.9$ \AA). The wavelength range for the fits was from 4750 to 6200 \AA, with selective masking of certain regions and lines susceptible to contamination by prominent sky lines or emission lines, as these could potentially skew the results. 

\begin{figure}
    \centering
    \includegraphics[scale=.5]{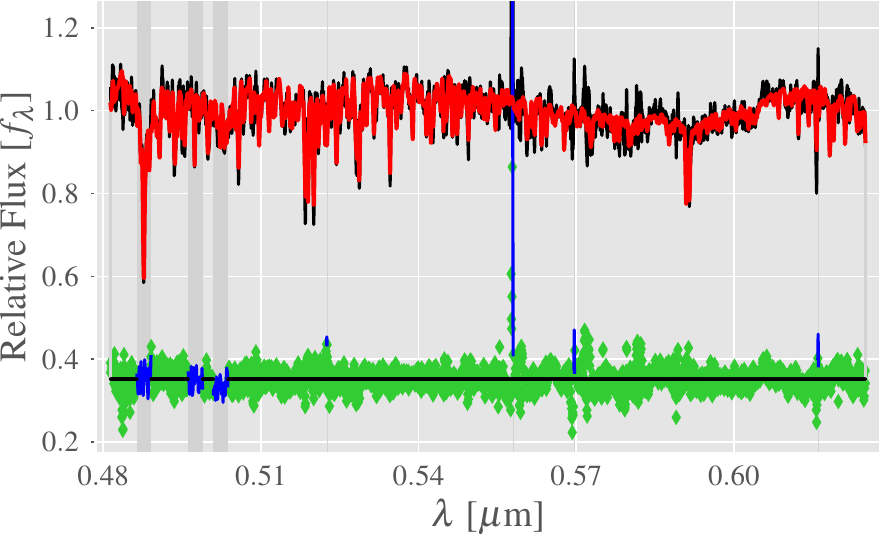}
    \caption{The spectrum of the central region of CGCG014-074 is shown in black. The red line represents the best fit and its residual is shown in green. Regions and lines discarded from the fit are shown in blue and light grey.}
    \label{fig:velocity_fit}
\end{figure}

Table \ref{tab:kinematics_populations} lists the apertures used at different galactocentric radii as well as the values for $V_\mathrm{rad}$ and $\sigma_*$, along with their respective uncertainties estimated by Monte Carlo simulations. In each simulation, each pixel of the spectra was resampled from a Gaussian distribution with a width equal to the observational error of that pixel. The parameter values and their associated uncertainties were obtained as the median and 1$\sigma$ estimation from $100$ simulations, respectively. This number of simulations is conventional for this type of analysis, as evidenced by previous studies \citep[e.g.,][]{Caceres2012, Salinas2012}.
Figure \ref{fig:velocity_fit} shows an example fit obtained for the spectrum in Table \ref{tab:kinematics_populations} corresponding to the central region of the galaxy.

\begin{table*}
	\centering
	\caption{Apertures used to extract the 1D spectra. The columns show the average galactocentric radii of each aperture and the signal-to-noise ratio measured at $4700$\AA. The spectrum corresponding to the central region of the galaxy is indicated in bold, while the negative and positive values of $r_g$ indicate apertures taken along the slit to the southeast and northeast, respectively. Columns 3 and 4 list the kinematic parameters (radial velocity and velocity dispersion) of CGCG014-074. Columns 5-8 list the values of the stellar population parameters (age and metallicity) weighted in luminosity and mass, respectively.}
	\label{tab:kinematics_populations}
    \begin{tabular}{cccccccc} 
        \hline
        \textit{r$_\mathrm{g}$}   & \textit{S/N} & V$_\mathrm{rad}$  &  $\sigma_*$ & \textit{Age$_\mathrm{light}$}  & [Z/H]$_\mathrm{light}$  &  \textit{Age$_\mathrm{mass}$} &  [Z/H]$_\mathrm{mass}$ \\
        \textit{(kpc/arcsec)} & \textit{(per \AA)}   &  \textit{(km\,s$^{-1}$)}  &  \textit{(km\,s$^{-1}$)} & \textit{(Gyr)}  &  \textit{(dex)}  &  \textit{(Gyr)}  &  \textit{(dex)} \\
        \hline
        -2.39\,/\,-35.7  & 6.4  &   985.3$\pm$9.1  & 24.3$\pm$11.0 & ---          &     ---        &     ---        &      ---         \\
        -1.01\,/\,-15.1  & 10.0 &   990.8$\pm$5.2  & 18.6$\pm$8.3 &  4.42$\pm$1.86 & -0.34$\pm$0.10 & 7.45$\pm$1.40 &  -0.29$\pm$0.08  \\
        -0.63\,/\,-9.52  & 19.0 &   990.1$\pm$2.4  & 17.2$\pm$5.3 &  5.91$\pm$1.13 & -0.48$\pm$0.06 & 8.31$\pm$1.23 &  -0.45$\pm$0.08  \\
        -0.36\,/\,-5.36  & 20.0 &   992.0$\pm$2.1  & 23.1$\pm$3.7 &  5.33$\pm$1.29 & -0.48$\pm$0.07 & 8.48$\pm$1.20 &  -0.39$\pm$0.08  \\
        -0.23\,/\,-3.44  & 20.4 &   994.8$\pm$1.9  & 21.6$\pm$2.9 &  2.85$\pm$1.45 & -0.39$\pm$0.18 & 4.00$\pm$1.41 &  -0.36$\pm$0.12  \\
        -0.13\,/\,-2.00  & 21.0 &   999.3$\pm$1.7  & 19.8$\pm$3.1 &  3.51$\pm$1.56 & -0.40$\pm$0.08 & 5.76$\pm$1.51 &  -0.54$\pm$0.06  \\
        -0.06\,/\,-0.96  & 22.0 &  1002.8$\pm$1.7  & 20.1$\pm$2.7 &  2.94$\pm$1.39 & -0.43$\pm$0.10 & 3.41$\pm$1.40 &  -0.50$\pm$0.11  \\
        ~\textbf{0.00}\,/\,~\textbf{0.00} & \textbf{22.5} &  \textbf{1005.5$\pm$1.6}  & \textbf{17.7$\pm$2.7} & \textbf{9.22$\pm$1.15} & \textbf{-0.87$\pm$0.30} & \textbf{9.40$\pm$1.14}  & \textbf{-0.81$\pm$0.27}    \\
        ~0.04\,/\,~0.64  & 21.1 &  1005.6$\pm$1.7  & 16.5$\pm$5.2 &  4.05$\pm$1.70 & -0.53$\pm$0.17 & 5.29$\pm$1.53 & -0.73$\pm$0.25   \\
        ~0.11\,/\,~1.68  & 21.5 &  1006.6$\pm$1.8  & 22.0$\pm$4.1 &  2.40$\pm$1.61 & -0.33$\pm$0.11 & 3.66$\pm$1.67 & -0.28$\pm$0.18   \\
        ~0.20\,/\,~2.96  & 19.2 &  1009.3$\pm$1.9  & 21.3$\pm$4.6 &  2.96$\pm$1.67 & -0.33$\pm$0.15 & 4.50$\pm$1.70 & -0.43$\pm$0.15   \\
        ~0.30\,/\,~4.56  & 19.7 &  1012.8$\pm$2.0  & 17.6$\pm$5.5 &  5.42$\pm$1.26 & -0.40$\pm$0.20 & 6.80$\pm$1.14 & -0.42$\pm$0.20   \\
        ~0.49\,/\,~7.28  & 19.3 &  1016.5$\pm$2.4  & 28.0$\pm$4.7 &  3.09$\pm$1.43 & -0.30$\pm$0.16 & 3.98$\pm$1.34 & -0.34$\pm$0.17   \\
        ~0.90\,/\,~13.5  & 10.1 &  1021.8$\pm$5.0  & 24.6$\pm$8.1 &  5.22$\pm$1.37 & -0.34$\pm$0.18 & 5.45$\pm$1.37 & -0.38$\pm$0.20   \\
        ~2.22\,/\,~33.1  & 7.3  &  1025.7$\pm$9.2  & 29.5$\pm$11.8 &   ---          &    ---         &   ---        &     ---          \\ 
        \hline
    \end{tabular}
\end{table*}

\begin{figure}
    \centering
    \includegraphics{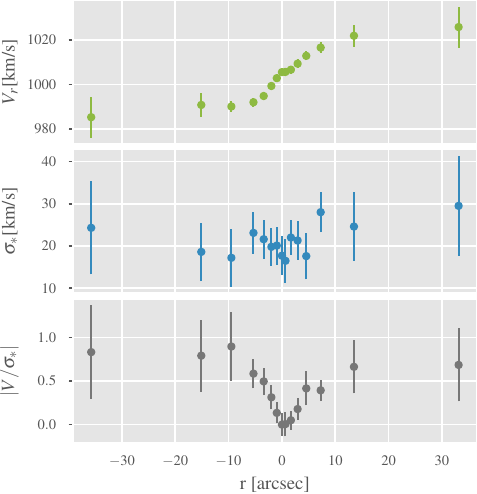} 
    \caption{Velocity profile of CGCG014-074. Top panel shows radial velocity versus galactocentric radii. Middle panel shows velocity dispersion versus galactocentric radii. Bottom panel shows the absolute value of the ratio $V/\sigma_*$ relative to the central velocity of the galaxy ($V_\mathrm{rad}$=1005.5 km/s) versus galactocentric radii.
 }
    \label{fig:kinematics}
\end{figure}

Figure \ref{fig:kinematics} shows the velocity profile of CGCG014-074 based on the values estimated by \textsc{pPXF}. As seen, there is strong evidence of rotation with an amplitude of $\sim20$ km/s. 
Notably, the $V/\sigma_*$ values show a progressive increase towards the outer regions, eventually reaching $V/\sigma_* \sim 1$. This observation lends substantial support to the plausible existence of a disk component within CGCG014-074. 
In addition, we calculate the anisotropy parameter $V_{max}/\sigma$ within an effective radius by correcting it for the inclination using the following expression $(V_{max}/\sigma)^* = (V_{max}/\sigma)\,[\varepsilon/(1-\varepsilon)]^{-1/2}$, being $\varepsilon$ the ellipticity. We estimate this parameter to determine whether the galaxy is pressure or rotationally-supported. To do this, we obtained the dispersion velocity by correcting the spectra in Table \ref{tab:kinematics_populations}, except for the outermost spectra, by shifting them to the same wavelength scale from the estimated radial velocity values. Once corrected, they were coadded into a single spectrum on which the dispersion velocity was obtained using \textsc{pPXF}. Regarding the maximum rotational velocity ($V_{max}$), we estimate the mean value from the two most distant estimated velocities on either side of the galactic centre but within the effective radius. Finally, the obtained value is $(V_{max}/\sigma)^* = 1.5 \pm 0.2$, indicating a rotationally supported system.

The central radial velocity value obtained for CGCG014-074 ($V_\mathrm{rad}^{cent}= 1005.5 \pm 1.6$ km/s) is in good agreement with the estimate made by \citet{Colless2003} (V$_\mathrm{hel}=998\pm54$ km/s) using data from the 2dF Galaxy Redshift Survey. Furthermore, the values of $V_\mathrm{rad}$ corresponding to the spectra of the central region, which includes the light from the compact object, indicate that it is part of the dwarf galaxy and constitutes its nucleus.

On the other hand, the velocity dispersion profile shows minimal, if any, discernible trend with radius, with a mean of $\langle \sigma_* \rangle=21.5$ km/s and a standard deviation of $3.9$ km/s. The $\sigma_*$ values obtained here are deemed reliable due to \textsc{pPXF}'s ability to accurately recover the velocity dispersion even below the spectral resolution \citep{Cappellari2017}, as in this case. The absence of a prominent central dip or peak in the $\sigma_*$ distribution suggests that the nucleus of CGCG014-074 exerts a negligible influence on its kinematics. 
In addition, this estimated $\langle \sigma_* \rangle$ value for the galaxy is consistent with those reported in the literature for early-type dwarf galaxies \citep[e.g.,][]{Chilingarian2009, Koleva2011, Forbes2011, Norris2014,Bidaran2020}.

\subsection{Mass estimates}
Based on the mean colour values obtained in Section \ref{sec:colourmaps}, we estimated the stellar mass of CGCG014-074 using the expressions from \citet{Zhang2017} that relate the colour of the galaxy to the mass-to-light ratio (M/L). We used the absolute solar magnitude in the $z$ filter of 4.50 \citep{Willmer2018} since it is less sensitive to metallicity variations. Considering an absolute magnitude of $M_z=-16.68$ mag for CGCG014-074, we obtained the value $M_*/L_z=1.134$ (with a dispersion of $0.102$). Consequently, the estimated stellar mass of CGCG014-074 is $3.3\times10^{8}$ M$_{\odot}$. This stellar mass value, together with the previously presented effective radius, is in the range exhibited by early-type dwarf galaxies \citep[e.g.,][]{Norris2014}.

In addition, from the determination of the kinematic parameters (see Section \ref{sec:kinematics}), we estimated the dynamical mass of CGCG014-074 following the guidelines of \citet{Toloba2009}. To do this, we define the total mass from the relation $M_{tot} = M_{press} + M_{rot}$, where $M_{press}$ is the mass obtained from the velocity dispersion, and $M_{rot}$ the mass inferred from the intrinsic rotation velocity of the galaxy.
Both terms are given by the expressions:

\begin{equation}\label{eq:masa_press}
M_{press} = \frac{C \: \sigma_*^2 \: R}{G} \, ,
\end{equation}
and
\begin{equation}\label{eq:masa_rot}
M_{rot} = \frac{R \: v_{max}^2}{G} \, ,
\end{equation}
where $C$ is the virial coefficient, $R$ is a measure of the system's size, $\sigma_*$ is the velocity dispersion of the system, $v_{max}$ the maximum rotational velocity, and $G$ is the gravitational constant.
To estimate the value of $C$, we used Equation 11 from \citet{Bertin2002}, which relates the value of the Sérsic index to this coefficient. In our case, considering the estimated mean value of $n=1.7$ ($C=7.59$) and the effective radius of 0.89 kpc, the total dynamical mass obtained for CGCG014-074 is $M_\mathrm{tot}=8.0 \times 10^{8} M_\odot$.

\subsection{Stellar Populations}
From the reduced 2D spectroscopic data (see Section \ref{sec:obs_data}), we extracted and flux-calibrated the spectrum of CGCG014-074 within an aperture of $192$ arcsec (12.8 kpc). This aperture was used to include the greatest amount of light from the galaxy. The integrated spectrum shows an SNR between 30 per \AA\ at 4800\,\AA\ to 50 per \AA\ at 6200\,\AA. We measure several Lick/IDS indices \citep{Worthey1997} within our spectral range. We use the $\chi^2$ minimization method of \citet{Proctor2004} together with the SSP models of \citet{Thomas2011} to determine the age, metallicity, and $\alpha$-element enhancement ratio [$\alpha$/Fe] of CGCG014-074. The SSP models were interpolated to obtain a smoother and finer grid of models in parameter space. 
Minimizing the $\chi^2$ distance between the indices Fe$_{4531}$, H$\beta$, Fe$_{5015}$, Mg$_2$, Mg$_b$, Fe$_{5270}$, Fe$_{5335}$ and Fe$_{5406}$ and the parameter grid \citep[e.g.,][]{Norris2006,Carlsten2017,Faifer2017}, we obtained the following light-weighted mean values for the integrated spectrum of the dwarf galaxy: 8.3$^{+1.4}_{-1.2}$ Gyr; $\mathrm{[Z/H]}=-0.59\pm0.11$ dex and [$\alpha$/Fe]$=0.08\pm0.06$ dex. The respective errors are derived from 100 Monte Carlo simulations of the data within the measured index errors. 

Subsequently, we used \textsc{pPXF} on the extracted spectrum to study the star formation history (SFH) of CGCG014-074. Unlike the kinematic analysis, we used the MILES\footnote{https://miles.iac.es} model library \citep{Vazdekis2015} for this purpose, which allows us to select a given initial mass function (IMF) and a specific $\alpha$-element abundance. In this case, we assume a Kroupa IMF and [$\alpha$/Fe]$=0.0$ dex since these are the MILES models closest to the value previously estimated with the LICK indices. The MILES models, with a resolution of 2.51 \AA, cover a wide range of ages ($0.03-14$ Gyr) and metallicities ($-2.27 <$ [Z/H] $< 0.4$ dex). During the spectral fitting, the science spectrum was reduced to the resolution of the templates and a multiplicative polynomial of order 10 was considered. In addition, \textsc{pPXF} was allowed to fit emission lines, if present. 

The obtained solutions were smoothed using the regularisation option to avoid spurious solutions and to reduce the degeneracy problem. To do this, it was sought that the solutions obtained had $\Delta \chi^2 \sim \sqrt{2N}$ \citep{Kacharov2018}, where $N$ is the number of pixels in the spectrum being fitted, and $\Delta \chi^2$ is the difference between the $\chi^2$ value of the current solution and that of the non-regularised case. This guarantees that the solution obtained is the smoothest one consistent with the observations.
We infer the uncertainties of the SSP grid weights through a bootstrap analysis, where the \textsc{pPXF} fit is repeated 100 times. This was done by resampling the pixels from the residual obtained from the unregularized best fit. This resampled spectrum is then used as input to \textsc{pPXF} taking into account a minimum regularization (R = 20). In this way, the variance of the weights is obtained solely from the variance of the spectrum, without forcing any prior regularisation.

When performing the fits, the SSP models are normalised in luminosity to a solar mass; thus, the solution provided by \textsc{pPXF} yields luminosity or mass weights for each fitted SSP, depending on the user's choice. From these weights, mass or luminosity-weighted averages of age and metallicity are derived using the following expressions:
\begin{equation}\label{eq:age}
\langle \log(Age)\rangle = \frac{\Sigma w_{i} \log(Age_{i})}{\Sigma w_{i}}
\end{equation}
\begin{equation}\label{eq:metallicity}
\langle \mathrm{[Z/H]} \rangle = \frac{\Sigma w_{i} \mathrm{[Z/H]}_{i}}{\Sigma w_{i}}
\end{equation}
where $w_{i}$ is the mass or luminosity fraction of a given model with age and metallicity.
Using these expressions, we calculate the average values of age and metallicity weighted in luminosity and mass for the integrated spectrum of the galaxy, obtaining: $5.3\pm1.1$ Gyr; $-0.62\pm0.05$ dex and $6.1\pm1.1$ Gyr, $-0.59\pm0.05$ dex, respectively.

\begin{figure}
    \centering
    \includegraphics[scale=.45]{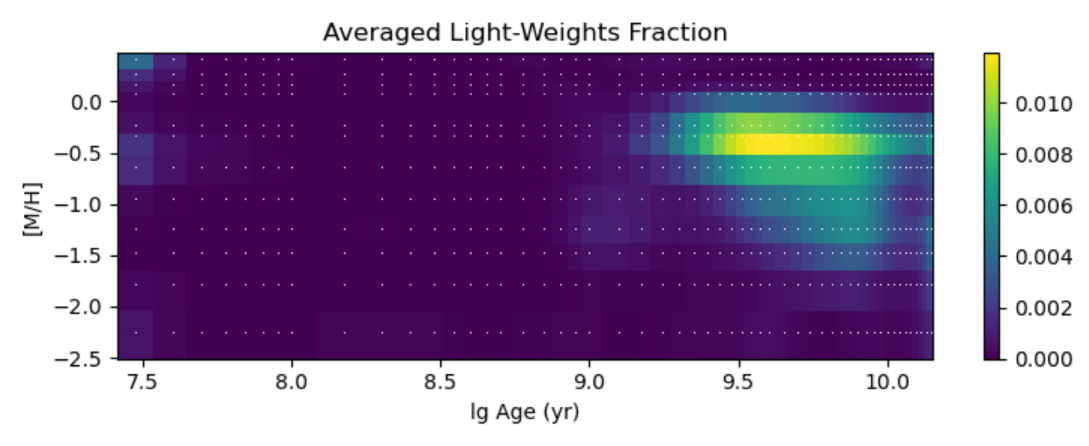} \\
    \includegraphics[scale=.45]{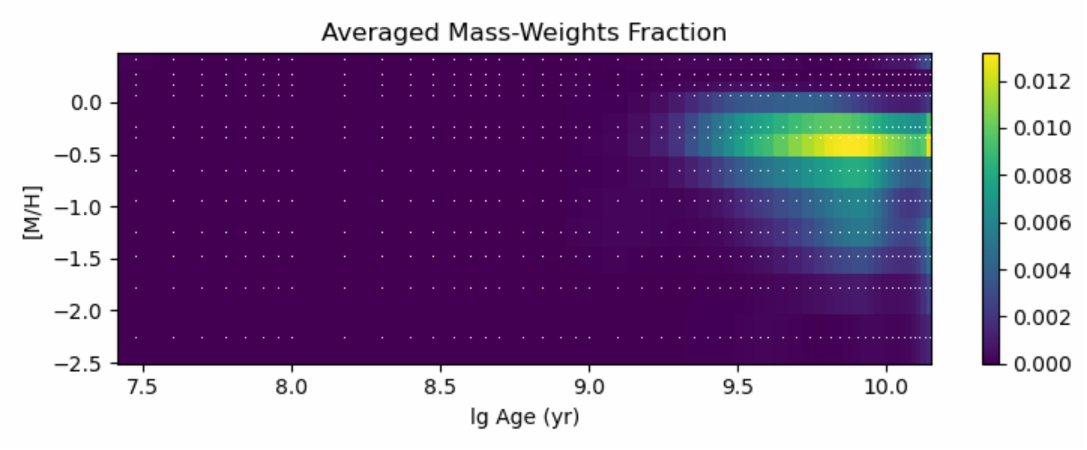} \\
    \includegraphics[scale=.45]{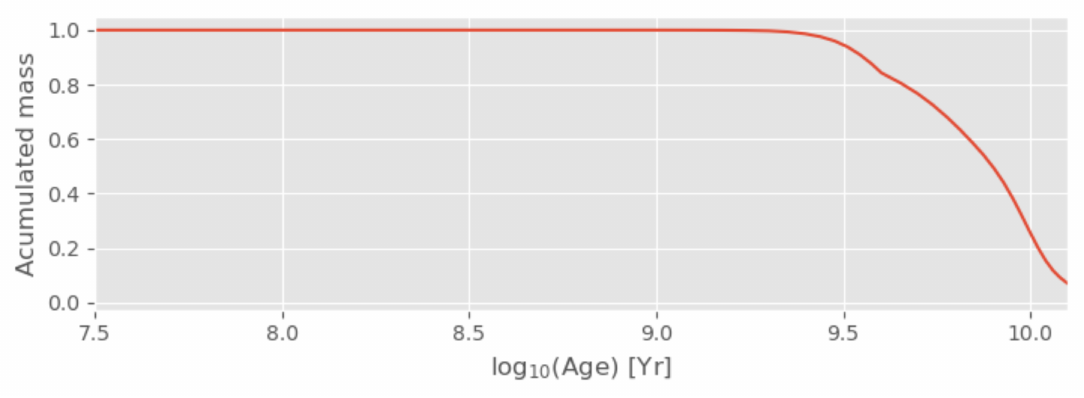}

    \caption{Output of the \textsc{pPXF} analysis using regularisation. 
    The top and middle panels show the SFHs weighted in light and mass, respectively. The lower panel shows the cumulative mass over time. }
    \label{fig:SFH}
\end{figure}

Figure \ref{fig:SFH} shows the luminosity and mass-weighted star formation histories (top and middle panels) and the cumulative mass as a function of age (bottom panel) for the integrated spectrum of CGCG014-074. Initially, we note that both SFHs behave similarly, showing the expected differences in the light contribution of stellar populations according to their ages. Specifically, CGCG014-074 appears to have experienced an extended SFH from its early stages ($\sim$13 Gyr) until about 2 Gyr ago, when it reached virtually all of its stellar mass. During this period, the metallicity shows a wide range of values from $\mathrm{[Z/H]}\sim-1.50$ dex to 0.0 dex. However, this range decreases towards younger ages, probably due to the enrichment of the interstellar medium and the new generation of more metal-rich stellar populations.

\begin{figure}
    \centering
    \includegraphics{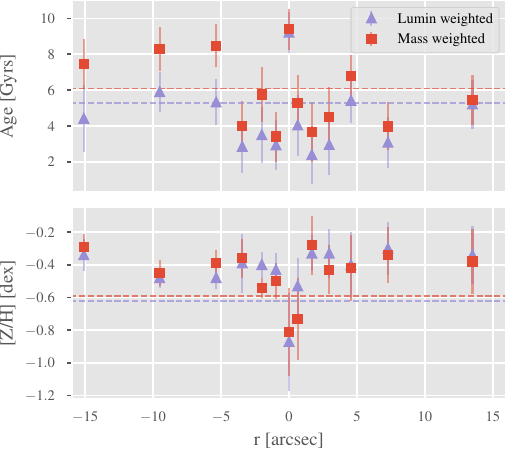}
    \caption{The age and metallicity weighted in light and mass (top and bottom panel, respectively), obtained from the best fits made with \textsc{pPXF}, for each spectrum at different galactocentric radii. Red and purple dashed lines indicate average values of age and metallicity weighted in luminosity and mass for the integrated spectrum of the galaxy. 
    }
    \label{fig:age_met_r}
\end{figure}

We ran \textsc{pPXF} again, but in this case on the individual spectra from Table \ref{tab:kinematics_populations}, except for those spectra furthest from the galactic centre due to their low SNR. We have used the expressions \ref{eq:age} and \ref{eq:metallicity} to obtain the mass- and light-weighted stellar population parameters for each of them. Columns $5-8$ of the same table list the values obtained for each spectrum, and Figure \ref{fig:age_met_r} shows the variation of these parameters as a function of the galactocentric radius. 

This figure shows that the central region corresponding to the nucleus of CGCG014-074 has an old age ($\sim$9.3 Gyr) and a low metallicity ($\mathrm{[Z/H]}\sim-0.84$ dex), both in its weighted values of luminosity and mass. Then, along the semi-major axis in the region corresponding to the disk of the galaxy, the age has approximately constant values ($\langle \mathrm{Age}_\mathrm{lum} \rangle  = 4.3$ Gyr; $\langle \mathrm{Age}_\mathrm{mass} \rangle = 5.7$ Gyr) in the range $1 < r_g < 6$ arcsec ($0.067 < r_g < 0.33$ kpc), increasing slightly outwards ($\langle \mathrm{Age}_\mathrm{lum} \rangle  = 4.6$ Gyr; $\langle \mathrm{Age}_\mathrm{mass} \rangle = 6.3$ Gyr). The metallicity, outside the nucleus and throughout the analysed region, remains relatively constant ($\langle \mathrm{[Z/H]_\mathrm{lum}} \rangle  = -0.39$ dex; $\langle \mathrm{[Z/H]_\mathrm{mass}} \rangle = -0.42$ dex).

\section{Summary and discussion}
\label{sec:summary}
This paper presents the photometric and spectroscopic analysis of the early-type dwarf galaxy CGCG014-074. The observations were obtained with the GMOS South instrument of the Gemini Observatory. The photometric data were obtained using the broadband filters $g'$, $r'$, $i'$ and $z'$, while the spectroscopic observations were made in the long-slit mode of the same instrument.

From the photometric analysis, we determined the surface brightness profiles of the galaxy and found that a simple single-component model cannot represent the light distribution of the galaxy. In this case, the best representation of these profiles was obtained using three components: a Gaussian profile describing the nucleus of CGCG014-074, and two Sérsic profiles, one representing a light excess in the innermost part of the galaxy ($0.2 \lesssim r_\mathrm{eq} \lesssim 0.8$ kpc) and the other for the outermost ($r_\mathrm{eq}\gtrsim 0.8$ kpc) extended stellar component. This result is consistent with the original classification of CGCG014-074 as a lenticular dwarf galaxy since such objects show surface brightness profiles of at least two components \citep{Sandage1984,Aguerri2005}. 

The innermost Sérsic profile in the four filters shows Sérsic index values of $n\sim0.8$. Considering that the value $n\sim 1$ represents an exponential disk, the result obtained here would indicate the presence of a stellar disk in CGCG014-074. These values are even comparable to those obtained for dwarf galaxies of equivalent magnitude located in clusters \citep[e.g.,][]{Janz2014,Venhola2019}.
In this sense, examination of the colour maps of the galaxy in these regions shows a smooth surface, with no visible evidence of dust and no colour differences between the inner disk and other components of the galaxy. From this analysis, we determined the absolute magnitudes for each filter (see table \ref{tab:ajustes_perfil_sersic}) and the effective radii (mean effective radius $\langle R_\mathrm{eff}\rangle=13.27\pm0.08$ arcsec; $0.89$ kpc), per filter, of CGCG014-074.

The isophotal analysis shows smooth changes in ellipticity and position angle (${\Delta \text{PA} = 13^\circ}$ and ${\Delta\varepsilon = 0.3}$) within $3 < r_\mathrm{eq} < 40$ arcsec ($0.2 < r_\mathrm{eq} < 2.8$ kpc). In the same galactocentric region, the kinematic study shows signs of rotation, with a velocity dispersion that does not vary with radius and has an average value of $\langle \sigma_* \rangle=21.5$ km/s. Furthermore, the anisotropy parameter of the galaxy results in $(V_{max}/\sigma)^* = 1.5 \pm 0.2$, indicating a rotationally supported system.
These results confirm the presence of a disk in CGCG014-074.
On the other hand, in the innermost region of the surface brightness profile, we observe significant variations in the isophotal parameters $\varepsilon$ and PA due to the presence of the nucleus, which has been confirmed kinematically. 

The classification of early-type dwarf galaxies has been a subject of debate due to the lack of a clear and consistent definition. Traditionally, these galaxies have been divided into dE (dwarf ellipticals) and dS0, though the distinction between the two remains ambiguous. Some studies, such as \citet{Lisker2006}, have challenged the existence of dS0 as a separate class, proposing instead that many of these galaxies might be disk systems rather than spheroids with an embedded disk component. With our validation, we can morphologically classify CGCG014-074 as a nucleated early-type dwarf galaxy featuring an embedded disk. Following the nomenclature suggested by \citeauthor{Lisker2006}, this galaxy would be classified as "dEdi,N", indicating its likely membership in a population of genuine disk galaxies.

Furthermore, the variation of the isophotal parameter shows that the major axis of the isophotes tends to point towards the dominant galaxy of the group (NGC\,4546) with increasing radii. Likewise, at $r_\mathrm{eq}\sim 25$ arcsec ($\sim 1.6$ kpc) the $B_4$ parameter changes from positive (disky isophotes) to negative (boxy isophotes). The occurrence of boxy isophotes in galaxies 
is usually associated with possible interaction or merger events \citep{Kormendy1996,Naab2006}.
The presence of this type of isophotes in CGCG014-074 could be explained as a consequence of the beginning of an interaction with the neighbouring giant galaxy NGC\,4546. Although the globular cluster system of NGC\,4546 and the surrounding region of CGCG014-074 was studied in \citet{Escudero2020}, no clustering or anisotropy of objects was initially found around the latter that could provide further evidence for this. In this same aspect, it is necessary to consider another potential object that could have influenced CGCG014-074 and that is the presence of the nucleated dwarf galaxy destroyed by tidal interaction with NGC\,4546 in the last $1-2$ Gyr \citep[NGC4546-UCD1; ][]{Norris2015}.

We studied the stellar population parameters of CGCG014-074 using two different methods. First, the values of age, metallicity and $\alpha$-element abundances were determined by analyzing the Lick/IDS indices on the total spectrum extracted from the dwarf galaxy. In this way, global parameters of CGCG014-074 were obtained, which are: 8.3$^{+1.4}_{-1.2}$ Gyr; $\mathrm{[Z/H]}=-0.59\pm0.11$ dex and [$\alpha$/Fe]$=0.08\pm0.06$ dex. This last value of [$\alpha$/Fe] was then used to run the full spectral fitting technique using \textsc{pPXF} to obtain its luminosity- and mass-weighted star formation history. This analysis revealed that CGCG014-074 underwent a prolonged period of stellar formation from its beginning ($\sim13$ Gyr) until about 2 Gyr ago when its star formation ceased and it reached 100\% of its stellar mass. In this age range, the metallicity covers a range of values from $\mathrm{[Z/H]} \sim -1.5$ to $\mathrm{[Z/H]} \sim 0.0$ dex, and then its dispersion decreases towards younger ages. It is necessary to emphasize that in principle it is not possible to determine whether this SFH was continuous or if it was due to different bursts of star formation. The average values of age and metallicity weighted in luminosity and mass obtained by \textsc{pPXF} yielded $5.3\pm1.1$ Gyr; $\mathrm{[Z/H]}=-0.62\pm0.05$ dex and $6.1\pm1.1$ Gyr, $\mathrm{[Z/H]}=-0.59\pm0.05$ dex, respectively, obtaining a difference compared to the Lick values. This discrepancy between the values obtained by the two methods has already been pointed out in other works in the literature \citep[e.g.,][]{Faifer2017,Carlsten2017}.

In the same context, observing the stellar parameters at different galactocentric radii, it is seen that the central region corresponding to the nucleus of CGCG014-074 has an older age ($\sim 9.3$ Gyr) compared to the regions that make up the stellar disk ($\sim 4.4$ Gyr). Likewise, the metallicity in the central region ($\mathrm{[Z/H]}\sim-0.84$ dex) is lower than in the regions external to it ($\mathrm{[Z/H]}\sim-0.40$ dex). This variation in age and metallicity in the galaxy is consistent with the observed flat photometric colour profile, indicating the effect of age-metallicity degeneration.
According to \citet{Fahrion2021}, these characteristics in CGCG014-074 suggest a formation scenario in which the nucleus primarily formed through the accretion of older, metal-poor globular clusters, while the younger, more metal-rich stellar disk indicates subsequent in situ star formation. This dual formation pathway is particularly relevant for galaxies with a mass around $M_{\text{gal}} \sim 10^9\,M_\odot$, where both globular cluster accretion and in situ star formation contribute to the build-up of the nuclear star cluster and the surrounding galaxy structure.

Finally, we have determined the total stellar and dynamical mass of CGCG014-074 by studying the photometric colour profiles and kinematic analysis, respectively. The values obtained are $M_\star= 3.3 \times 10^8 M_\odot$ and $M_{tot} = 8.0 \times 10^8 M_\odot$, and they are in good agreement with the values obtained for other early-type dwarf galaxies, \citep[e.g.,][]{Norris2014,Eftekhari2022}.

As for the possible origin of CGCG014-074, several works using simulations \citep{Naab2006,Robertson2006,Hoffman2010} have suggested that large mergers of metal-rich disk galaxies could produce slow-rotating early-type galaxies ($V/\sigma_* \sim 1$), with the possibility of a kinematically decoupled core \citep{Hoffman2010}. In these cases, the galaxy's gas would be driven towards the centre of the forming galaxy, creating an internal disk with a scale of $1 R_\mathrm{eff}$, while the stars at larger galactocentric radii would undergo dry mergers, giving rise boxy isophotes in these regions. Although low-mass galaxies are expected to undergo fewer mergers, this scenario is supported by several observations, and many simulations suggest that this mechanism of formation could be possible in a low-density environment \citep{Bekki2015,Graham2012,Pak2016,Watts2016}. If this were the case for CGCG014-074, we would expect the inner disk of the galaxy to be younger than the outer regions.

Another possibility is that isolated low-mass galaxies, which have stopped forming stars and have not experienced dramatic events such as major mergers or tidal shocks, are just the building blocks for the formation of hierarchical structures. These objects could form disks and steadily acquire angular momentum through the accumulation of gas from the cosmic web \citep{Maccio2006}, and through minor mergers.

In this context, CGCG014-074 displays characteristics aligning with both scenarios discussed earlier. On the one hand, the galaxy shows a rotating inner disk, an extended stellar formation \citep[as is the case for late-type galaxies;][]{Boselli2001}, with the cessation of activity $\sim 2$ Gyr ago, and boxy isophotes towards the outer regions. Initially, no kinematically decoupled core is detected, nor are there any signs or traces of a major merger. On the other hand, the possible accretion of primordial gas from the environment in the recent past does not seem to be the case either, as the galaxy's metallicity is not notably low. As shown in this study, the disk of the galaxy has a higher metallicity than its nucleus, possibly indicating a secular evolution in CGCG014-074. Together, these features position CGCG014-074 as a probable building block galaxy that has evolved passively throughout its history.


\section*{Acknowledgements}
We thank the referee, Professor Dr. Reynier Peletier, for his constructive comments and suggestions that helped improve this paper. 

This work was funded with grants from Consejo Nacional de Investigaciones Cientificas y Tecnicas de la Republica Argentina, and Universidad Nacional de La Plata (Argentina). 

Based on observations obtained at the international Gemini Observatory, a program of NSF’s NOIRLab, which is managed by the Association of Universities for Research in Astronomy (AURA) under a cooperative agreement with the National Science Foundation on behalf of the Gemini Observatory partnership: the National Science Foundation (United States), National Research Council (Canada), Agencia Nacional de Investigaci\'{o}n y Desarrollo (Chile), Ministerio de Ciencia, Tecnolog\'{i}a e Innovaci\'{o}n (Argentina), Minist\'{e}rio da Ci\^{e}ncia, Tecnologia, Inova\c{c}\~{o}es e Comunica\c{c}\~{o}es (Brazil), and Korea Astronomy and Space Science Institute (Republic of Korea). The Gemini program ID are GS-2014A-Q-30 and GS-2020A-Q-130.

The DESI Legacy Imaging Surveys consist of three individual and complementary projects: the Dark Energy Camera Legacy Survey (DECaLS), the Beijing-Arizona Sky Survey (BASS), and the Mayall z-band Legacy Survey (MzLS). DECaLS, BASS and MzLS together include data obtained, respectively, at the Blanco telescope, Cerro Tololo Inter-American Observatory, NSF’s NOIRLab; the Bok telescope, Steward Observatory, University of Arizona; and the Mayall telescope, Kitt Peak National Observatory, NOIRLab. NOIRLab is operated by the Association of Universities for Research in Astronomy (AURA) under a cooperative agreement with the National Science Foundation. Pipeline processing and analyses of the data were supported by NOIRLab and the Lawrence Berkeley National Laboratory (LBNL). Legacy Surveys also uses data products from the Near-Earth Object Wide-field Infrared Survey Explorer (NEOWISE), a project of the Jet Propulsion Laboratory/California Institute of Technology, funded by the National Aeronautics and Space Administration. Legacy Surveys was supported by: the Director, Office of Science, Office of High Energy Physics of the U.S. Department of Energy; the National Energy Research Scientific Computing Center, a DOE Office of Science User Facility; the U.S. National Science Foundation, Division of Astronomical Sciences; the National Astronomical Observatories of China, the Chinese Academy of Sciences and the Chinese National Natural Science Foundation. LBNL is managed by the Regents of the University of California under contract to the U.S. Department of Energy.

This research has made use of the NASA/IPAC Extragalactic Database (NED),
which is operated by the Jet Propulsion Laboratory, California Institute of Technology,
under contract with the National Aeronautics and Space Administration.


\section*{Data Availability}

The data underlying this article are available in the Gemini Observatory Archive at https://archive.gemini.edu/searchform, and can be accessed with Program ID:GS-2014A-Q-30/PI: Escudero, C. and  Program ID:GS-2020A-Q-130/PI: Escudero, C.



\bibliographystyle{mnras}
\bibliography{mnras_template} 








\bsp	
\label{lastpage}
\end{document}